\documentclass[a4paper,fleqn,usenatbib]{mnras}

\usepackage[T1]{fontenc}
\usepackage{ae,aecompl}

\usepackage{graphicx}
\usepackage{color}
\usepackage{amssymb}
\usepackage{amsmath}
\usepackage{multicol}
\usepackage{multirow}
\input{colordvi.tex}

\newcommand{\kvec}{{\boldsymbol{k}}}

\title{Constraining the EoR model parameters with the 21cm bispectrum}
\author[H.Shimabukuro et al.]
  {Hayato Shimabukuro$^1$$^{,2}$$^{,3}$,
  Shintaro Yoshiura$^2$, Keitaro Takahashi $^2$, Shuichiro Yokoyama$^{4}$
  \newauthor
  and Kiyotomo Ichiki$^1$$^{,5}$\\
  $^1$Department of Physics, Graduate School of Science, 
Nagoya University, Aichi, 464-8602, Japan\\
 $^2$Department of Physics, Kumamoto University, Kumamoto, Japan\\
  $^3$Observatoire de Paris, LERMA, Paris, France\\
  $^4$Department of Physics, Rikkyo University, Tokyo, Japan\\
   $^5$Kobayashi-Maskawa Institute for the Origin of Particles and the Universe, Nagoya University, Nagoya, 464-8602, Japan
   }

\begin{document}

\label{firstpage}
\pagerange{\pageref{firstpage}--\pageref{lastpage}}
\maketitle

\begin{abstract}
We perform a Fisher analysis to estimate the expected constraints on the Epoch of Reionization (EoR) model parameters (i.e., minimum virial temperature, the ionizing efficiency and the mean free path of ionizing photons) taking into account the thermal noise of the existing telescopes, MWA and LOFAR.  We consider how the inclusion of the 21cm bispectrum improves the constraints compared to using the power spectrum alone.  Assuming that we perfectly remove the foregrounds, we found that the bispectrum, which is calculated by the 21cmFAST code, can constrain the EoR model parameters more tightly than the power spectrum since the bispectrum is more sensitive to the EoR model parameters than the power spectrum.  We also found that degeneracy between the EoR model parameters can be reduced by combining the bispectrum with the power spectrum.  

\end{abstract}

RUP-16-23

\section{Introduction}

Following the cosmic recombination, there were no luminous objects.  This epoch is often called the `Dark Ages (DA)'.  The DA ended with the formation of the first luminous objects \cite{yos,Fialkov:2013uwm,Visbal:2012aw,2011A&A...527A..93S}, this epoch is called the `Cosmic Dawn (CD)'.  According to the standard hierarchical structure formation model based on Lambda CDM ($\Lambda \rm CDM$) cosmology, massive objects such as galaxies formed after the formation of smaller objects.   These first generation objects played an important role in both thermal and ionization histories of the intergalactic medium (IGM)  \cite{Mesinger:2012ys, Christian:2013gma,Fialkov:2014kta,2016arXiv160204407Y}.  As structure formation proceeded, ionizing photons emitted by galaxies resulted in the ionization of the IGM.  This transition in the state of hydrogen in the IGM is known as cosmic reionization.  This phase change is called `Epoch of Reionization'.  Several recent observations have provided us with useful information on the EoR.  For example, high-$z$ QSO absorption lines imprinted in their spectra indicate that reionization concluded by $z\sim 6$(e.g.  \cite{Fan:2006dp}), and the rapid evolution of the Ly-$\alpha$ luminosity function with redshift constrains the neutral hydrogen fraction at $z>6$ \cite{2014ApJ...797...16K}.  Furthermore, the optical depth of Thomson scattering has been measured and the value obtained by PLANCK is $\tau_{e}=0.066\pm0.016$  \cite{2015arXiv150702704P}.  This implies that reionization occurs at $z=8.9^{+2.5}_{-2.0}$ with instantaneous reionization scenario.  

The redshifted 21cm line signal, which is the emission due to hyperfine structure in a neutral hydrogen atom, is expected to be a promising tool to probe matter density fluctuations, the ionization state and the spin temperature at the EoR \cite{fur, 2012RPPh...75h6901P}. Although the cosmological 21cm signal has not been observed yet, it is hoped that improvements to current instruments and foreground removal methods push observational 21cm cosmology into a new era in the near future.  In addition, on-going projects such as MWA \cite{Tingay:2012ps}, LOFAR \cite{Rottgering:2003jh} and PAPER \cite{2015ApJ...801...51J} have the potential to statistically detect the 21cm signal.  Furthermore, a future instrument, like the SKA\cite{2013ExA....36..235M} and HERA\cite{2016arXiv160607473D}, should be able to detect the 21cm power spectrum at higher redshifts beyond the EoR and to map the brightness temperature  \cite{Mesinger:2013nua, 2015aska.confE..12P, 2015aska.confE..11M, 2016arXiv160301961H}.

One of the fundamental statistical measures of the 21cm signal is the 21cm power spectrum.  If the 21cm signal follows a Gaussian distribution, we can extract all of the statistical information from the power spectrum alone.  However, we expect the 21cm signal to have  non-Gaussian features in its distribution due to inhomogeneous astrophysical processes, such as X-ray heating and reionization \cite{2007MNRAS.379.1647W,2008MNRAS.384.1069B,2010MNRAS.406.2521I,2016MNRAS.458.3003S}.  In order to estimate the non-Gaussian features of the 21cm signal, various statistical quantities have been suggested such as the bispectrum, one-point statistics, and Minkowski functionals \cite{Watkinson:2013fea,2015MNRAS.451.4986S,2015arXiv150507108W,2016MNRAS.458.3003S,2016arXiv160202351Y}.  In our previous works, we examined the properties of the 21cm bispectrum and its detectability \cite{2015MNRAS.451.4785Y, 2016MNRAS.458.3003S}.  Yoshiura et al, calculated the sensitivity of the 21cm bispectrum by estimating the contribution from the thermal noise.  They showed that the 21cm bispectrum at the EoR could be detected by MWA and LOFAR at large scales $k\lesssim 0.3{\rm Mpc}^{-1}$ and at $k\lesssim 0.7$ with SKA \cite{2015MNRAS.451.4785Y}.  
As observational techniques are further developed, we expect to obtain much more information on the EoR from combinations of the 21cm power spectrum and the bispectrum.  It is imperative to estimate the parameters of EoR models if we are to succeed in detecting the cosmic 21cm signal.  Several previous works have  performed forecasts for the EoR parameters by using the Markov Chain Monte Carlo (MCMC) method or a Fisher forecast applied to the 21cm power spectrum or the global 21cm signal \cite{2014ApJ...782...66P, 2015MNRAS.449.4246G, 2015arXiv151008815L, 2015ApJ...813...11M, 2015arXiv151000271H}.  In addition, in a previous study, Kubota et al. performed a Fisher analysis on the variance and skewness of the brightness temperature \cite{2016arXiv160202873K}. 

In this paper, we consider a forecast of the parameter constraints by using a Fisher analysis of on-going observations of the 21cm signal.  We focus on MWA and LOFAR as first generation instruments.  A previous work performed a Fisher analysis on the 21cm power spectrum to constrain the EoR model parameters with MWA and LOFAR\cite{2014ApJ...782...66P} and obtained 1 $\sigma$ errors of $10-20\%$ for the fiducial value of the EoR model parameters. In our work, we estimate how these constraints on the EoR parameters can be improved by including the 21cm bispectrum.

In this paper, we employ the best fit values of the standard cosmological parameters obtained in  \cite{Komatsu:2010fb}.

\section{Formulation and set up}

\subsection{Formulation for the 21cm bispectrum}

A fundamental quantity of the 21cm signal is the differential brightness temperature, which is described as the spin temperature offset from the CMB temperature given by (see, e.g,  \cite{fur})
\begin{align}
\delta T_{b}(\nu) &= \frac{T_{{\rm S}}-T_{\gamma}}{1+z}(1-e^{-\tau_{\nu_{0}}})  \nonumber \\
                  &\quad \sim 27x_{{\rm H}}(1+\delta_{m})\bigg(\frac{H}{dv_{r}/dr+H}\bigg)\bigg(1-\frac{T_{\gamma}}{T_{{\rm S}}}\bigg) \nonumber \\
                  &\quad \times \bigg(\frac{1+z}{10}\frac{0.15}{\Omega_{m}h^{2}}\bigg)^{1/2}\bigg(\frac{\Omega_{b}h^{2}}{0.023}\bigg) [{\rm mK}].
\label{eq:brightness}
\end{align}
Here, $T_{\rm S}$ and $T_{\gamma}$ respectively represent the gas spin temperature and the CMB temperature, $\tau_{\nu_{0}}$ is the optical depth in the 21cm rest frame frequency $\nu_{0} = 1420.4~{\rm MHz}$, $x_{\rm H}$ is the neutral fraction of the hydrogen gas, $\delta_{m}({\bold x},z) \equiv \rho/\bar{\rho} -1$ is the evolved matter overdensity, $H(z)$ is the Hubble parameter and $dv_{r}/dr$ is the comoving gradient of the gas velocity along the line of sight.  All quantities are evaluated at a redshift of $z = \nu_{0}/\nu - 1$.

Let us focus on the spatial distribution of the brightness temperature. The spatial fluctuation of the brightness temperature  can be defined as
\begin{equation}
\delta_{21}({\bold x}) \equiv (\delta T_b({\bold x}) - \langle \delta T_b \rangle)/ \langle \delta T_b \rangle,
\end{equation}
where $\langle \delta T_b \rangle$ is the mean brightness temperature obtained from the brightness temperature map and $\langle ...\rangle$ expresses the ensemble average. From this definition, we have the power spectrum of $\delta_{21}$ defined as
\begin{equation}
\langle \delta_{21}({\bold k}) \delta_{21}({\bold k^{\prime}})\rangle
= (2\pi)^3 \delta({\bold k}+{\bold k^{\prime}}) P_{21}({\bold k}),
\label{eq:ps_def}
\end{equation}
If the statistics of the brightness temperature fluctuations is a pure Gaussian, the statistical information of the brightness temperature should be completely characterized by the power spectrum. The statistics of the brightness temperature fluctuations completely follows that of the density fluctuations $\delta_m$ if both of the spin temperature and the neutral fraction are completely homogeneous. However, in the CD and EoR eras, the spin temperature and the neutral fraction were spatially inhomogeneous and the statistics of the spatial fluctuations of those quantities would be highly non-Gaussian due to the various astrophysical effects. Accordingly, the statistics of the brightness temperature fluctuations would deviate from the pure Gaussian and it will be important to investigate the non-Gaussian features of the brightness temperature fluctuations.  Such a non-Gaussian feature can be investigated through the skewness of the one-point distribution functions as done in one of the our previous works \cite{2015MNRAS.451.4986S}.  However, the scale-dependent feature is integrated out in the skewness. Conversely, the higher order correlation functions in Fourier space such as the bispectrum and the trispectrum characterize the non-Gaussian features and contain the scale-dependent information. Here, in order to determine the non-Gaussian features of the brightness temperature fluctuations $\delta_{21}$, we focus on the bispectrum of $\delta_{21}$ which is given by
\begin{equation}
\langle \delta_{21}({\bold k_{1}}) \delta_{21}({\bold k_{2}}) \delta_{21}({\bold k_{3}})\rangle = (2 \pi)^3 \delta({\bold k_{1}}+{\bold k_{2}}+{\bold k_{3}})B({\bold k_{1}},{\bold k_{2}},{\bold k_{3}}).
\label{eq:bs_def}
\end{equation}
In \cite{2016MNRAS.458.3003S}, we used absolute value of bispectrum as an estimator. Thus, that bispectrum estimator includes non-zero value of imaginary part of the bispectrum although they should be zero \cite{2008PhRvD..78b3523S.  The reason why the imaginary part of the bispectrum did not becomes zero in Shimabukuro et al.2016 was that the imaginary part was not canceled because they only measured the positive half of $k$-space.  In this work, we instead use real part of the bispectrum as an estimator although statistical fluctuations are reduced.}  In order to calculate the bispectrum, we need to characterize the shape of the bispectrum in $k$-space.  In this work, we choose the equilateral type bispectrum ($k_1 = k_2 =k_3=k$) because the equilateral type of bispectrum normalized by wavenumber shows a stronger signal than other configurations [see  \cite{2016MNRAS.458.3003S}].  



\subsection{Calculation of the 21cm bispectrum}

We calculate the bispectrum of the brightness temperature fluctuations by making use of 21cmFAST  \cite{Mesinger:2007pd,2011MNRAS.411..955M}. This code is based on a semi-numerical model of reionization and the thermal history of the IGM, and generates maps of matter density, velocity, spin temperature, ionized fraction and brightness temperature at the designated redshifts. 

We perform simulations in a $200 {\rm Mpc}^3$ comoving box with $300^3$ grids, which corresponds to 0.66 comoving {\rm Mpc} resolution or $\sim$ 14.1 arcsec and a $1.19 {\rm deg}^{2}$ field of view at 127 {\rm MHz} (${\it z}$ = 10), from $z = 200$ to $z = 5$ adopting the following fiducial parameter set, $(\zeta, \zeta_{X}, T_{\rm vir}, R_{\rm mfp}) = (15, 10^{56}/M_{\odot}, 10^4~{\rm K}, 30~{\rm Mpc})$.  Here, $\zeta$ is the ionizing efficiency, $\zeta_{X}$ is the number of X-ray photons emitted by the source per solar mass, $T_{\rm vir}$ is the minimum virial temperature of halos which produce ionizing photons, and $R_{\rm mfp}$ is the mean free path of ionizing photons through the IGM. In our calculation, for simplicity, we ignore, the gradient of peculiar velocity whose contribution to the brightness temperature is relatively small (a few \%)  \cite{Ghara:2014yfa}.  We performed 10 realization calculations for every parameter set and then take the average bispectrum from these.

\subsection{Parameter dependence of the 21cm bispectrum}

We study the parameter dependence of the 21~cm bispectrum in order to prepare for the Fisher forecast.  We choose three key parameters as the EoR model parameters. We briefly summarize the key parameters below: \\

1. $\zeta$, {\it the ionizing} {\it efficiency}: $\zeta$ is composed of a number of parameters related to ionizing photons escaping from high redshift galaxies and given as $\zeta=f_{\rm esc}f_{*}N_{\gamma}/(1+\overline{n}_{\rm rec})$ \cite{fur}.  Here, $f_{\rm esc}$ is the fraction of ionizing photons escaping from galaxies into the IGM, $f_{*}$ is the fraction converted from baryons to stars, $N_{\gamma}$ is the number of ionizing photons per baryon in stars and $\overline{n}_{\rm rec}$ is the mean recombination rate per baryon.  In our calculation, we adopt $\zeta=15$ as the fiducial value to satisfy observed constraints on the ionization history.  \\

2. $T_{{\rm vir}}$, {\it the minimum} {\it virial} {\it temperature} {\it of} {\it halos} {\it producing} {\it ionizing} {\it photons}:  $T_{{\rm vir}}$ parameterizes the minimum mass of halos producing ionizing photons at the EoR.  Typically, $T_{\rm vir}$ is chosen to be $10^{4} {\rm K}$ corresponding to the temperature above which atomic cooling becomes effective.  $T_{\rm vir}$ includes the physics of high redshift galaxy formation.  If there is no radiative feedback, atomic cooling is thought to become effective at $T_{\rm vir}$=$10^{4}{\rm K}$.  Hydrogen molecule cooling becomes effective below this temperature.  If stars, or star forming galaxies, begin to form in a halo and radiative feedback by such objects exists, the minimum virial temperature is expected to become higher since radiative feedback, such as the photodissociation of $\rm H_{2}$, prevents the gas from cooling \cite{2013MNRAS.432.3340S}.  Conversely, positive feedback, such as the enhancement of $\rm H_{2}$ molecules due to the increase of electrons, pushes the minimum virial temperature to lower values because cooling by molecular hydrogen becomes more effective.  We parameterize $T_{\rm vir}$ as the parameter responsible for uncertainties in the radiative feedback effects discussed above.\\

3. $R_{\rm mfp}$, {\it the maximum mean free path of ionizing photons}: This parameter determines the maximum HII bubble size.  Physically, the mean free path of the ionizing photons is determined by the number density and the optical depth of Lyman-limit systems.  In our calculation, we choose $R_{\rm mfp}$=30[\rm comoving Mpc] as the fiducial value.\\

\begin{figure*}
\centering
\vspace{-25mm}
\includegraphics[width=1.0\hsize]{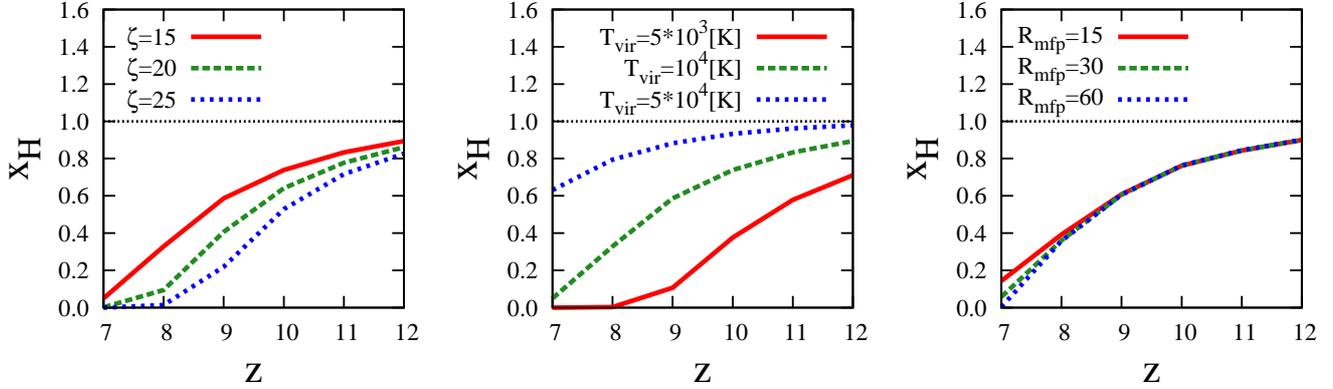}
\vspace{-35mm}
\caption{Ionization histories with varying $\zeta$({\it left}),  $T_{{\rm vir}}$({\it middle}) and $R_{\rm mfp}$({\it right}).  We adopt $\zeta=15, 20$ and 25, $T_{{\rm vir}}$=$5\times10^{3}, 10^{4}$ and $5\times 10^{4}[{\rm K}]$ and $R_{\rm mfp}=15, 30$ and 60 [Mpc]. }
\label{fig:ionization}
\end{figure*}

\begin{figure*}
\centering
\includegraphics[width=1\hsize]{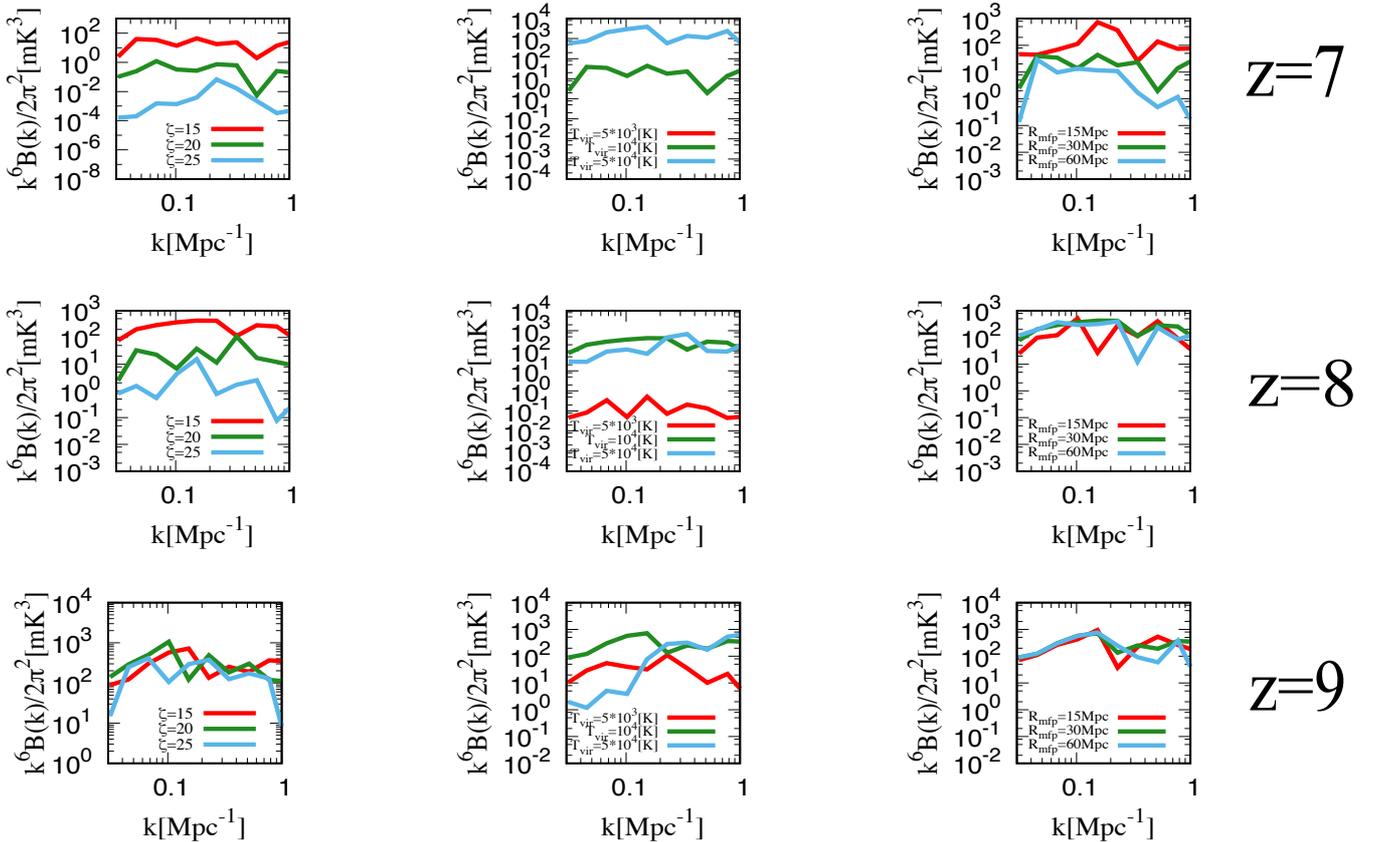}
\caption{The 21cm bispectrum as a function of wavenumber at $z=7$ ({\it 1st row}), 8({\it 2nd row}), \rm and 9({\it 3rd row}) with varying $\zeta$({\it left}), $T_{{\rm vir}}$({\it middle}) and $R_{\rm mfp}$({\it right}).  We adopt $\zeta=15, 20,$ \rm and 25 and $T_{{\rm vir}}=5\times 10^{3}, 10^{4}$ and $5\times 10^{4} [{\rm K}]$, $R_{\rm mfp}=15, 30$, \rm and $60$ [Mpc].  }
\label{fig:bs_params1}
\end{figure*}

We show the parameter dependence of the ionization history in Fig. \ref{fig:ionization}.  For illustrative purposes, we adopt $\zeta$=15, 20 and 25 (because reionization would end later than $z=6$ which clashes with our constraints from quasars etc, we do not go lower than the fiducial value),  $T_{{\rm vir}}=5\times 10^{3} \rm K, 10^{4} \rm K, $ and $5\times 10^{4} \rm K$ and $R_{\rm mfp}$=15 \rm Mpc, 30 \rm Mpc and 60 \rm Mpc.   Fig.\ref{fig:ionization} shows that larger $\zeta$ and smaller $T_{\rm vir}$ cause earlier reionization.  Larger $\zeta$ means that much more photons can contribute to the ionization of the neutral hydrogen gas.  This leads to faster progression of the EoR.  Since lower $T_{{\rm vir}}$ corresponds to a smaller halo mass, the formation epoch of halos capable of producing ionizing photons shifts to earlier times.  This is because larger $\zeta$ and smaller $T_{\rm vir}$ cause earlier reionization.  Furthermore, larger $R_{\rm mfp}$ causes efficient reionization because the large mean free path of the ionizing photons can result in large ionized bubbles.  However, the ionization history does not depend on $R_{\rm mfp}$ at higher redshifts when reionization did not progress efficiently.  This is because $R_{\rm mfp}$ affects the epoch after the ionized bubbles have grown to some extent.  In Fig.\ref{fig:bs_params1}, we show the scale dependence of the bispectrum at each redshift for various $\zeta$, $T_{\rm vir}$ and $R_{\rm map}$. In the first column of this figure, we can see that larger $\zeta$ suppresses the bispectrum as the redshift decreases.  This is because higher $\zeta$ drives reionization to earlier times and the neutral fraction becomes smaller overall.  In the second column, we can see that the power of the bispectrum is suppressed in the case of smaller $T_{\rm vir}$ at lower redshift, especially at larger scales.  Smaller  $T_{\rm vir}$ results in large numbers of haloes which are capable of producing ionizing photons.  These haloes produce large numbers of ionized bubbles which grow as reionization proceeds and then the 21cm signal coming from these bubbles becomes small.  Therefore, the power of the bispectrum corresponding to the bubble size is suppressed.  At $z=7$, reionization has finished in the case of $T_{\rm vir}=5\times 10^{3}[\rm K]$ and thus the 21cm signal becomes zero.  As we can see from Fig. \ref{fig:ionization}, the effect of $R_{\rm mfp}$ appears in the epoch after the neutral hydrogen fraction decreases to less than $\sim 0.5$ (lower redshift).  Therefore, the effect of $R_{\rm mfp}$ on the bispectrum is also slight at $z$=9 and 8 and only becomes apparent at $z=7$.


%

\section{Fisher information matrix \& estimation of thermal noise for the bispectrum}

In order to forecast constraints on the EoR model parameters, we use the Fisher information matrix ${\bf F}_{ij}$.  Given the observational data, the maximum likelihood analysis gives a set of parameters which maximize the likelihood function ${\cal L}$ (the probability distribution function for the measured data set as a function of the model parameters).  The Fisher formalism assumes that the likelihood function ${\cal L}$ is a multi-dimensional Gaussian of the given parameters.  Using the Fisher analysis  \cite{2009arXiv0906.4123C, 2010LNP...800..147V}, we can estimate the forecast errors on the model parameters with the supposed instruments. 

%
%
%
%
%

The Fisher matrices for the 21~cm power spectrum and the 21~cm bispectrum are respectively given by

\begin{eqnarray}
\label{eq:fisher_PS_BS}
{\bf F}_{ij,{\rm PS}}=\sum_{l}^{N} \left(\frac{1}{\delta P_{N}(k_{l})}\right)^{2}\frac{\partial P(k_l;\vec{p})}{\partial p_{i}}\frac{\partial P(k_l;\vec{p})}{\partial p_{j}}\bigg|_{\vec{p}=\vec{p}_{\rm fid}}\\
{\bf F}_{ij,{\rm BS}}=\sum_{l}^{N} \left(\frac{1}{\delta B_{N}(k_{l})}\right)^{2}\frac{\partial B(k_l;\vec{p})}{\partial p_{i}}\frac{\partial B(k_l;\vec{p})}{\partial p_{j}}\bigg|_{\vec{p}=\vec{p}_{\rm fid}}
\end{eqnarray}
where $\vec{p}$ is the model parameter vector, $\vec{p}=(p_{1}, p_{2},\cdot \cdot \cdot)$ and $\vec{p}_{\rm fid}$ is a set of fiducial model parameters, $\vec{p}_{\rm fid}=(p_{1,{\rm fid}}, p_{2, {\rm fid}},\cdot \cdot \cdot)$.  $l$ expresses the $l$-th bin of the wavenumber.  $\delta P_{N},\delta B_{N}$ are thermal noise of the power spectrum and the bispectrum, respectively.  We calculate the derivative of both power spectrum and bispectrum from brightness temperature map obtained from 21cmFAST.  As described in 2.2, we also calculate derivative from 10 realizations simulation and then take the average from these.  We perform numerical derivative with $d\zeta=0.05$, $d T_{\rm vir}$=100, $d R_{\rm mfp}=0.05$.

Note that we need to take the error covariance into account for precise evaluations of the Fisher matrix. Previous work shows that off-diagonal terms of error covariance of 21cm power spectrum have statistically non-zero values and thus this implies different wavenumber is independent, especially at smaller scales ($k\gtrsim 0.6{\rm Mpc^{-1}}$) \cite{2016MNRAS.456.1936M}.  We expect that error covariance of the bispectrum is also important.  However, we ignore this for simplification and because we are focusing on the relatively large scales accessible by the MWA and LOFAR ($k\lesssim 0.3 {\rm Mpc^{-1}}$).  The evaluation of error covariance of the 21cm bispectrum is our future work.

Given the Fisher matrix, we can estimate the expected 1-$\sigma$ error of the $i$-$th$ parameter:

\begin{equation}
\label{eq:1sigma}
\sigma_{p_{i}}=\sqrt{{\bf F^{-1}}_{ii}}.
\end{equation}



\begin{figure*}
\centering
\vspace{-20mm}
\includegraphics[width=1.0\hsize]{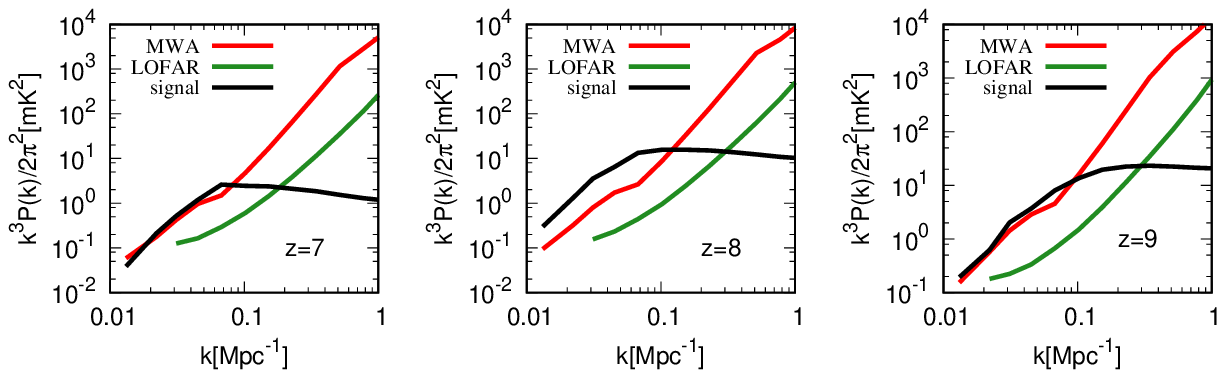}
\vspace{-60mm}
\vspace{-15mm}
\includegraphics[width=1.0\hsize]{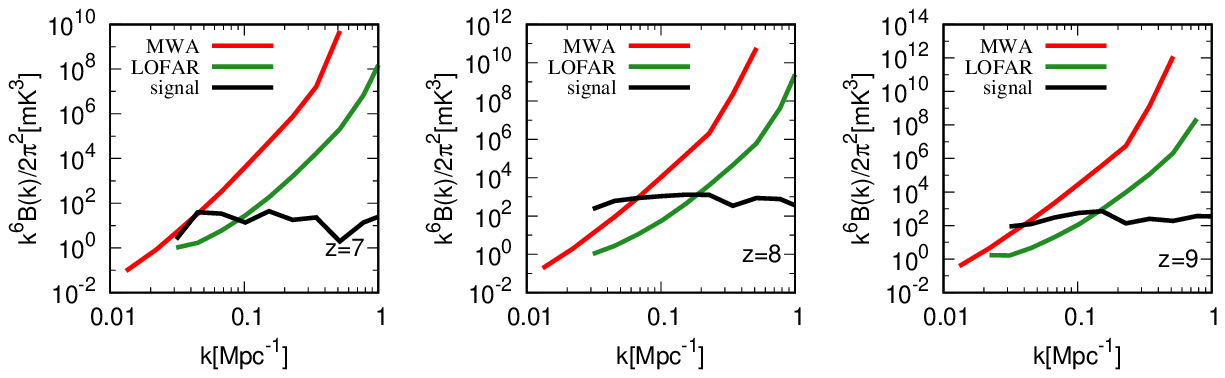}
\vspace{-20mm}
\caption{Comparison of 21cm power spectrum signal ({\it top}) and bispectrum({\it bottom}) (both are represented by short-dashed line) with thermal noise for various telescopes at $z$=7,8,9.  As telescopes, we choose the MWA(solid line), LOFAR(long-dashed line). }  
\label{fig:SN}
\end{figure*}

Next, we estimate the thermal noises of the power spectrum and bispectrum.  As opposed to the power spectrum of thermal noise, the ensemble average of the bispectrum of the thermal noise is actually zero if the thermal noise follows a Gaussian distribution.  However, the variance of the thermal noise bispectrum is non zero and this variance contributes to the 21cm bispectrum signal. The 1$\sigma$ error on the 21cm bispectrum due to thermal noise was derived by  \cite{2015MNRAS.451.4785Y}.  We use the formula for the spherically-averaged power spectrum of the thermal noise and the variance of the spherically-averaged thermal noise bispectrum shown in eqs.\ref{eq:dPk}, \ref{eq:dBk} respectively, which were derived by \cite{2006ApJ...653..815M, 2015MNRAS.451.4785Y}. 

\begin{align}
\delta P_N(k) \approx &\nonumber \bigg[ k^3 \int_{\arccos[\min(\frac{y k}{2 \pi},1)]}^{\arcsin[\min(\frac{k_*}{k}, 1)]} d\theta \sin{\theta}\\
&\times \frac{\epsilon (n(k \sin{\theta}))^2 A_e^3 B^2 t_0^2}{(2 \pi)^2 x^2 y \lambda^6 T_{\rm sys}^4} \bigg]^{-1/2},
\label{eq:dPk}
\end{align}

\begin{eqnarray}
\label{eq:dBk}
\delta B_{N}(k)
&=& \frac{(2\pi)^{\frac{5}{2}}}{\sqrt{\Delta \theta_2} k^{5/2} \epsilon}
    \left(\frac{x^2y \lambda^2}{A_e}\right)
    \left(\frac{T^2_{\rm sys} \lambda^2}{A_e B t_0}\right)^{\frac{3}{2}} \nonumber\\
  &&  \times \bigg[\int {d\theta_1} \int {d\alpha} \sin{\theta_1} \sin{\theta_2}\sin{\gamma(\theta_1,\alpha)} \nonumber\\
   &&            ~ n(\kvec_1) n(\kvec_2) n(\kvec_3)\bigg]^{-\frac{1}{2}},
\end{eqnarray}
where $k_{*}$ is the largest transverse wavenumber vector corresponding to the maximum baseline length.  The lower limit of the integral is determined by the pixel size.  The other quantities are the wavenumber $\lambda$, the system temperature $T_{\rm sys}$, the effective area $A_{e}$, the bandwidth $B$, the integral time $t_{0}$ and the number density of the baselines $n$. $x$ and $y$ , which are determined by the assumed cosmology, are the quantities which convert $uv$ space to Fourier space $\vec{k}$.  Other quantities ($\theta, \alpha$) are angles in Fourier space.  Please refer to \cite{2006ApJ...653..815M, 2015MNRAS.451.4785Y} for a detailed explanation.  Here, we use the telescope parameters listed in table 1 of  \cite{2015MNRAS.451.4785Y} (although they assume a MWA with 512 tiles (MWA-512T), we assume a MWA with 256 tiles (MWA-256T)  in order to reflect the instrument being constructed. Thus, we reduce the number of antennae by half in this work).


In Fig.\ref{fig:SN}, we show the scale dependence of the power spectrum and the bispectrum (as in Fig.\ref{fig:bs_params1}, we plot real part of the bispectrum), respectively, with thermal noise estimated for MWA and LOFAR. We assumed a total observation time of 1000 hours.  Note that we multiply the square (cube) of the average brightness temperature for the power spectrum (bispectrum) to simulate the observation of the 21 cm brightness temperature.  This is because an interferometer is not sensitive to the average signal. From this figure, we can see that the noise increases at smaller scales in both bispectrum and power spectrum.  This is because the number of longer baselines corresponding to smaller scales is deficient.  Conversely, the sensitivity at large scales is limited by the field of view.

For both bispectrum and power spectrum noises, we cannot calculate the sensitivity at $k\lesssim 0.03{\rm Mpc}^{-1}$ for the LOFAR telescope because the scales $k\lesssim 0.03{\rm Mpc}^{-1}$ are beyond the field of view of LOFAR.  Comparing the sensitivity of the power spectrum with that for the bispectrum, the signal to noise ratio in the case of the power spectrum is slightly larger than that in the case of the bispectrum.  If we focus just on the sensitivity, the 21~cm power spectrum is more detectable than the bispectrum.  However, the estimation of the expected constraint does not depend only on the sensitivity but also on the parameter dependences in the power spectrum or bispectrum.

\section{Result}
We now show the result of the Fisher analysis applied to the power spectrum and the bispectrum.  Here, we focus on the equilateral type bispectrum.  As previous mentioned, we constrain the EoR model parameters using current telescopes, MWA and LOFAR, to study how the bispectrum improves the constraint on the EoR parameters. Note that we use both power spectrum and bispectrum for the Fisher analysis for $k$=0.03 -1.0 ${\rm Mpc}^{-1}$ divided into 9 bins.

We show the confidence regions of the EoR model parameters.  Note that the confidence regions obtained by the Fisher analysis include physically meaningless regions such as $T_{\rm vir}<0$.  Thus, we put a physically meaningful boundary condition on the parameter space and exclude the negative value regions. 

\begin{figure*}
\centering
\vspace{+30mm}
\includegraphics[width=0.8\hsize,bb=0 0 480 190]{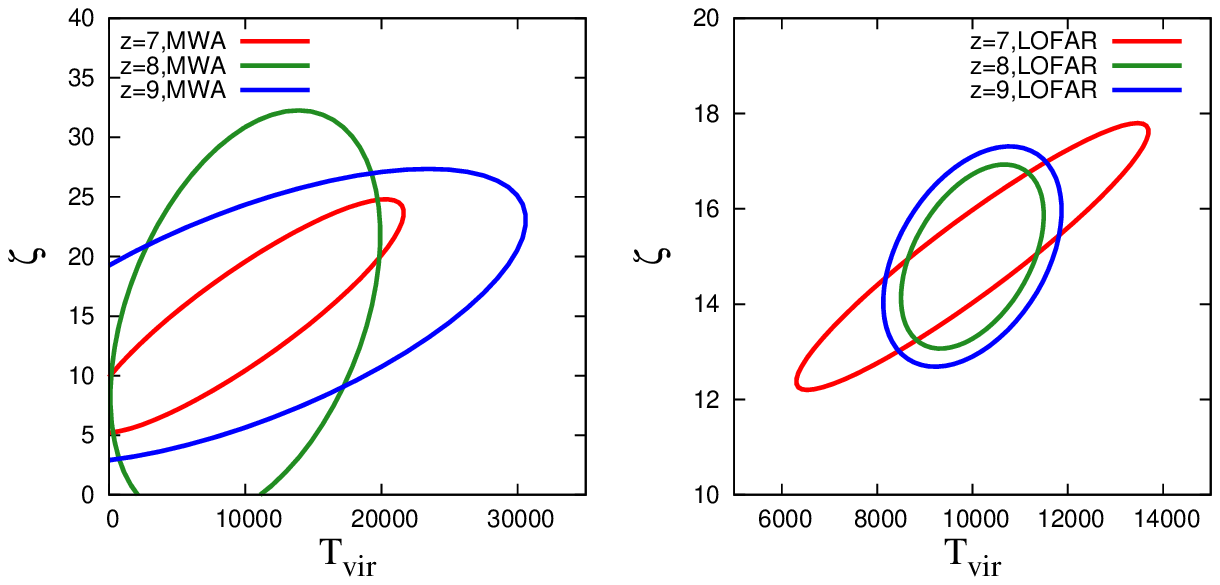}
\includegraphics[width=0.8\hsize,bb=0 0 480 190]{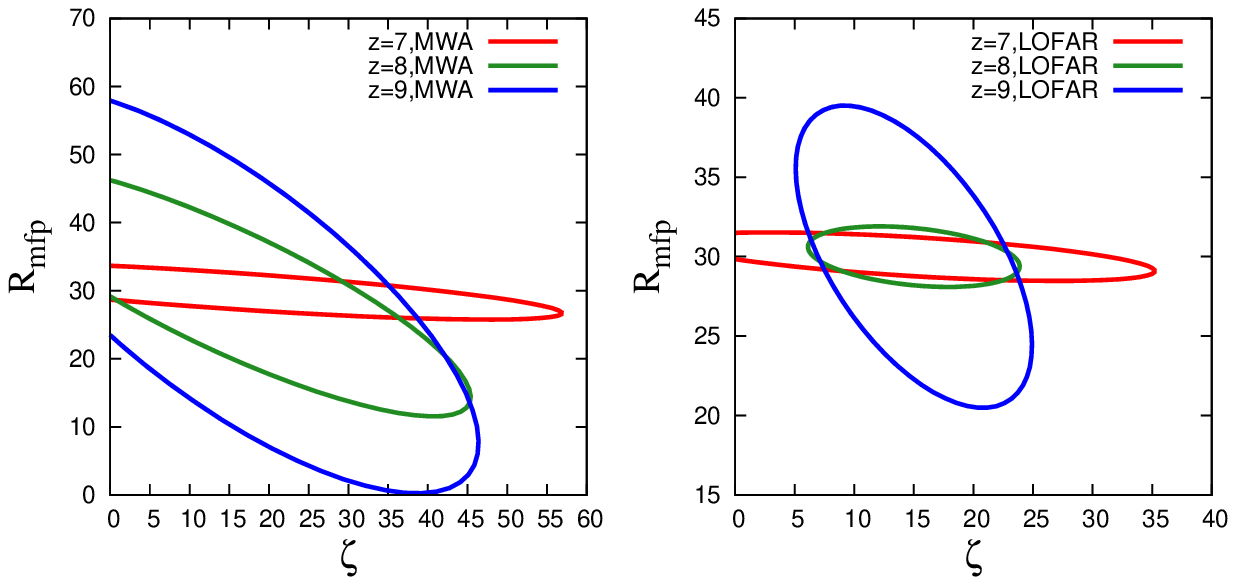}
\includegraphics[width=0.8\hsize,bb=0 0 480 190]{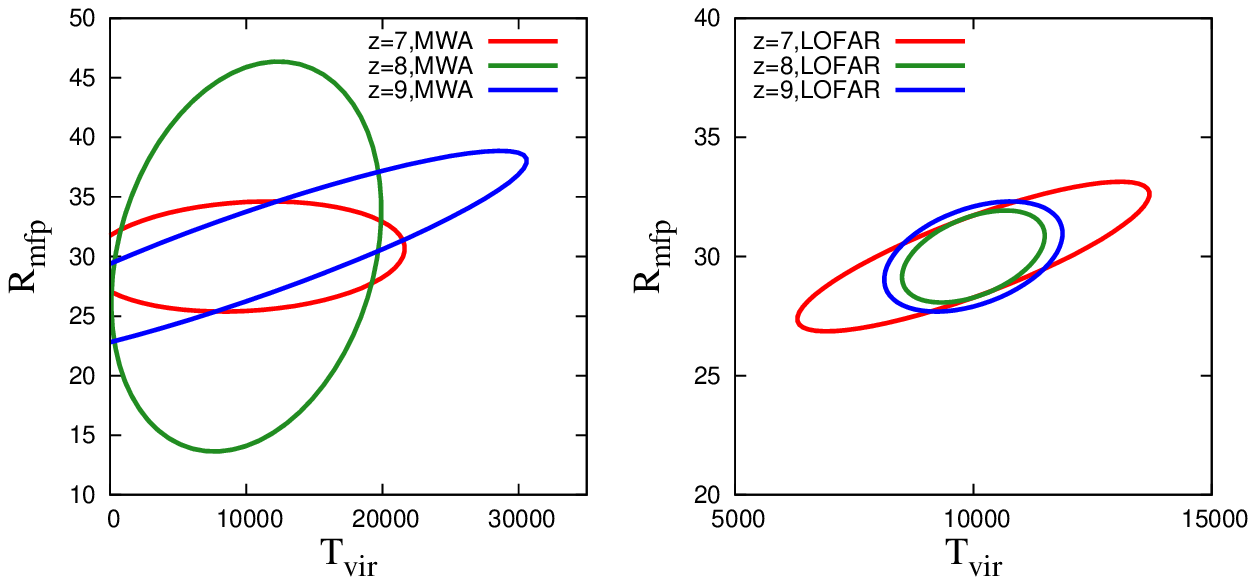}
\vspace{-16mm}
\caption{1-$\sigma$ contours of the EoR model parameters obtained by the bispectrum assuming MWA({\it left}) and LOFAR({\it right}).  }
\label{fig:parameter1}
\end{figure*}

First, we show constraints on the EoR model parameters obtained by the bispectrum at $z$=7, 8 and 9 in Fig.\ref{fig:parameter1}.  We can see that the constraints at $z=$8 are stronger than that those at other redshifts.  This is because the 21cm bispectrum as a function of the redshift has a peak at $z\sim 8$ in our model and therefore the bispectrum is most sensitive to the EoR parameters at $z\sim 8$.  We can also see that the constraint obtained assuming LOFAR is tighter than that obtained assuming MWA.  This is because the sensitivity and resolution of LOFAR are better than those of MWA.  The physical meaning of the inclination of the ellipse is as follows.   If $\zeta$ and $R_{\rm mfp}$ become larger, neutral hydrogen atoms are ionized more efficiently.  Similarly, decreasing $T_{\rm vir }$ also drives the progression of the reionization.  Both increasing $\zeta$ and $R_{\rm mfp}$ and decreasing $T_{\rm vir}$ play the same role in reionization.  Thus, there is a degeneracy between these parameters. 

\begin{figure*}
\centering
\vspace{+30mm}
\includegraphics[width=0.8\hsize,bb=0 0 480 190]{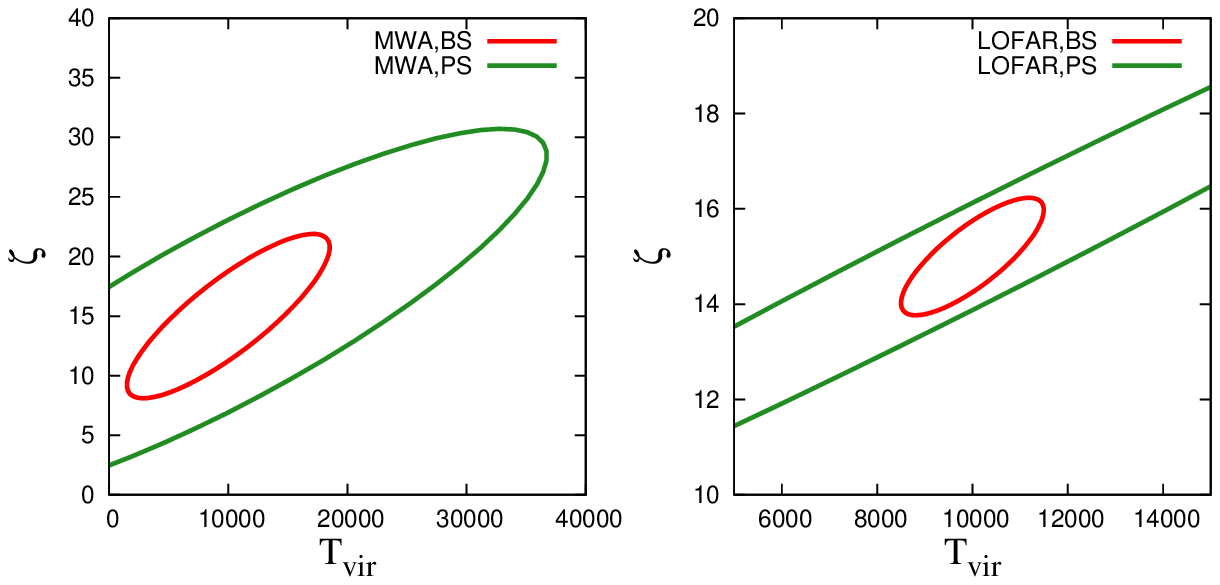}
\includegraphics[width=0.8\hsize,bb=0 0 480 190]{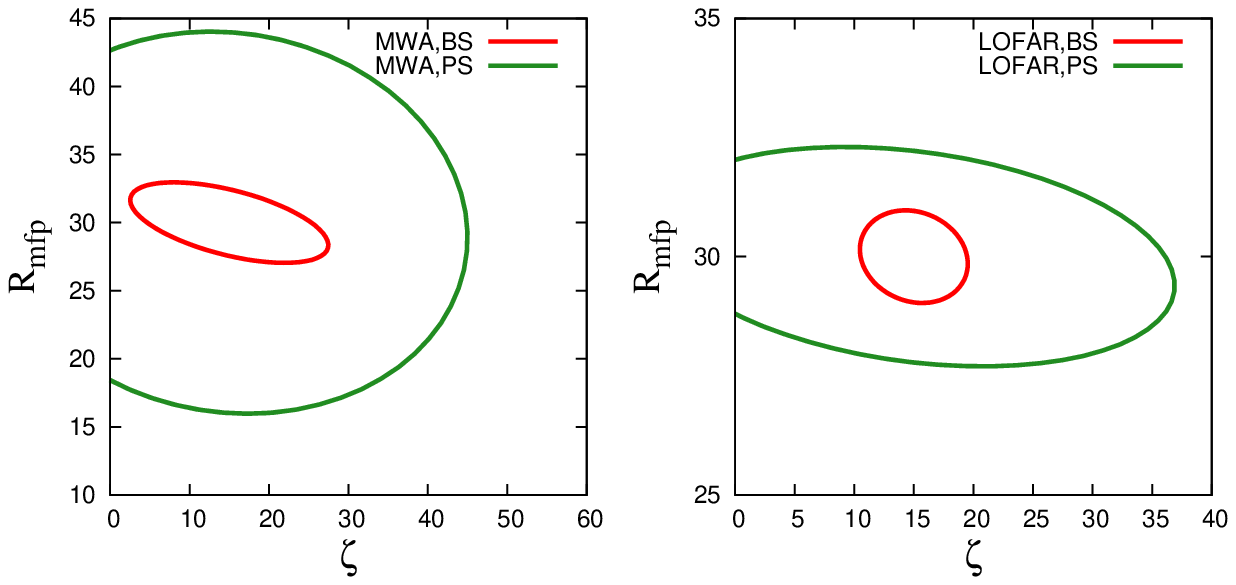}
\includegraphics[width=0.8\hsize,bb=0 0 480 190]{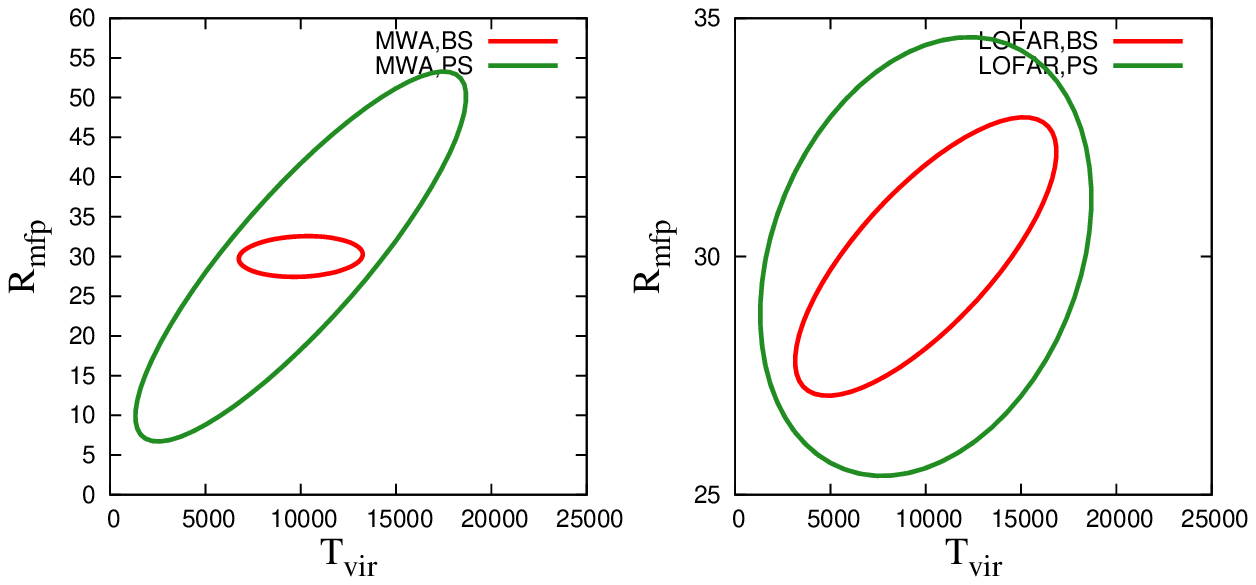}
\vspace{-16mm}
\caption{1-$\sigma$ contours of the EoR model parameters obtained by the bispectrum (red) and power spectrum(green) assuming MWA({\it left}) and LOFAR({\it right}).  Here we use the bispectrum and the power spectrum at $z$=7, 8 and 9.}
\label{fig:parameter_z}
\end{figure*}

Next, we compare the size of the constraints obtained by the power spectrum with that obtained by the bispectrum.  We show the result in Fig.\ref{fig:parameter_z} and table. \ref{table:fisher2}.  Here, we combine the bispectrum and the power spectrum at $z$=7, 8 and 9.  We find that the constraints from the bispectrum are tighter than those from the power spectrum.  As you can see table. \ref{table:fisher2}, if we use the bispectrum, each parameter can be determined with an accuracy $\sim 90$ percent for MWA and within $\sim 16$ percent for LOFAR.  This accuracy is 1-2 orders of the magnitude better than constraints obtained from the power spectrum.  We also find that using the LOFAR telescope results in tighter constraints than using the MWA telescopes.

\begin{figure*}
\centering
\vspace{-30mm}
\includegraphics[width=1.0\hsize]{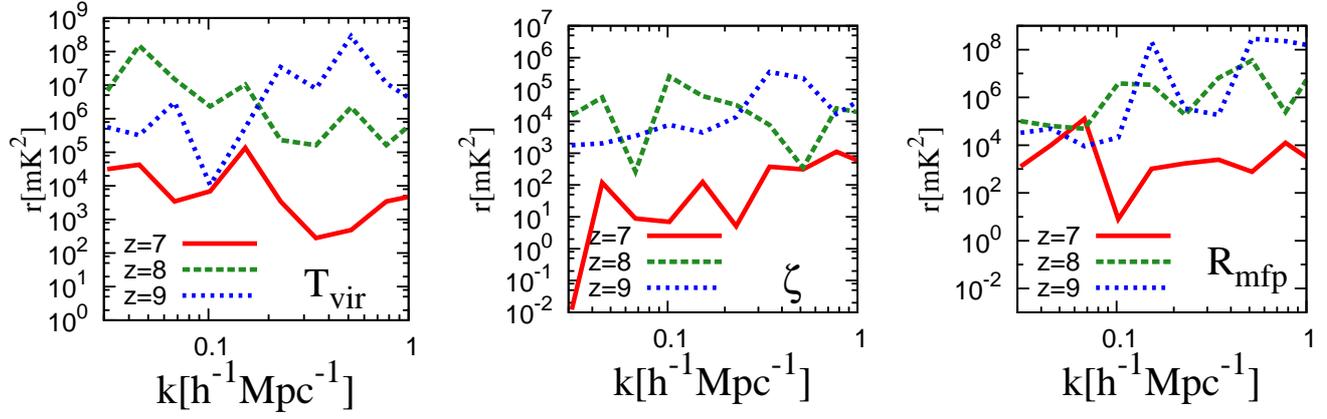}
\vspace{-16mm}
\caption{Ratio of the square of the derivative of the bispectrum and the power spectrum with respect to the virial temperature ({\it left}), ionizing efficiency ({\it middle}) and maximum mean free path ({\it left}).  $r=\left(\frac{\partial B}{\partial p_{i}}/\frac{\partial P}{\partial p_{i}}\right)^{2}$. } 
\label{fig:ratio_derivative}
\end{figure*}

We find that the bispectrum can constrain the EoR parameters tighter than the power spectrum although the signal to noise ratio of the power spectrum is better than that of the bispectrum.  The bispectrum can give tighter constraints because the derivative of the bispectrum with respect to the EoR parameters is much larger than that of the power spectrum.  However, we know that the Fisher matrix is determined not only by the derivative of the signal but also by the thermal noise.  In order to study the balance between the derivative of the signal and the thermal noise, we show the ratio of the square of the derivative with respect to the virial temperature, the ionizing efficiency and the maximum mean free path for the bispectrum and the power spectrum, $r=\left(\frac{\partial B}{\partial p_{i}}/\frac{\partial P}{\partial p_{i}}\right)^{2}$ in Fig.\ref{fig:ratio_derivative}.  we show this as functions of the wave number.   The figure implies that the derivative of the bispectrum with respect to parameters is larger than that of the power spectrum ($r>1$).  Consequently, the derivative of the bispectrum contributes to the Fisher matrix more than that of the power spectrum (a larger Fisher matrix results in tighter constraints).  The difference between the redshifts is remarkable for the ratio of the derivative.  In particular, the ratio of the derivative is large at $z$=8,9.

\begin{table}
\centering
\label{my-label}
\begin{tabular}{lllll}
   & $\Delta\zeta_{\rm fid}$/$\zeta_{\rm fid}$ & $\Delta T_{\rm vir}$/$T_{\rm vir, fid}$ & $\Delta R_{\rm mfp}$/$R_{\rm mfp,fid}$& \\ \hline
PS, MWA &  2.33   & 2.89   & 0.633   &  \\ \hline
PS, LOFAR & 0.971   & 0.996    & 0.338    &  \\ \hline
BS, MWA   &   0.600   &   0.915  &   0.452  &  \\ \hline
BS, LOFAR  & 0.157  & 0.0289  & 0.134 & \\ \hline
  \end{tabular}
 \caption{Constraints on $\zeta$, ${\rm T_{vir}}$ and $R_{\rm mfp}$ estimated using the Fisher forecast with the power spectrum(PS) and the bispectrum(BS) at  $z$=7, 8, 9.}
\label{table:fisher2}
\end{table}


%
%
%

\section{Summary \& Discussion}
In order to explore the EoR parameter region with MWA and LOFAR observations, we estimated the expected 1-$\sigma$ errors and constrained the parameter region using a Fisher analysis with the 21~cm power spectrum and the bispectrum.  First, we found that we can put tighter constraints on the EoR parameters with LOFAR than with MWA.  LOFAR can give 1-2 orders of magnitude better constraints on the parameters than MWA because the thermal noise for LOFAR is lower than that for MWA.  The difference in the specifications between MWA and LOFAR comes from the effective area and the maximum baseline length.  Although the number density of the antennae in the core region of MWA is larger, the larger effective area of LOFAR compensates for its smaller number density of antennae.


Next, we found that the forecast errors obtained by the bispectrum are better than those obtained by the power spectrum.  The bispectrum can give constraints on each parameter with an accuracy within $\sim 3 - $90 $\%$ although the power spectrum can constrain each parameter within 30 - 290 $\%$.  This is because the derivative of the bispectrum is more sensitive to the EoR parameters than that of the power spectrum and the ratio of the derivative to thermal noise for the bispectrum is larger than that of the power spectrum.  We also found that the combination of the power spectrum and bispectrum error contours reduces the degeneracy between EoR parameters as shown in Fig.\ref{fig:parameter_z}, in particular for LOFAR.  Therefore, we expect that we can obtain tight constraints on the EoR parameters by combining the bispectrum and the power spectrum. 

What we have to address is that Fisher matrix is simply not a good approximation for first generation instruments because the likelihood of the EoR parameters for those instruments has non-gaussianity as shown in Fig.5 of \cite{2015MNRAS.449.4246G}.  Furthermore, the Fisher forecast we have done would underestimate the error contours than their work partly because we ignore the error covariance.  Thus, it is our future work to evaluate the effect of error covariance.  However, we have shown the superiority of the bispectrum for constraints of  the EoR parameters compared with power spectrum under the same condition.  We believe that this is remarkable point of our work.

\section*{Acknowledgement}
This work is supported by Grant-in-Aid from the Ministry of Education, Culture, Sports, Science and Technology (MEXT) of Japan, Nos. 24340048(K.T. and K.I.), 26610048, 15H05896, 16H05999. (K.T.), 25-3015(H.S.), 16J01585(Yoshiura) and 15K17659,16H01103(S.Y.).


\label{lastpage}

\end{document}